\documentclass[a4paper,fleqn,usenatbib]{mnras}

\usepackage{newtxtext,newtxmath}
\usepackage{multirow}
\usepackage{bm}
\usepackage{color}

\usepackage[T1]{fontenc}
\usepackage{ae,aecompl}

\usepackage{graphicx}	
\usepackage{amsmath}	
\usepackage{amssymb}	

\title[Galaxy-Quasar Strong Lens System]{SDSS\,J1640+1932: a spectacular galaxy-quasar strong lens system}

\author[Lin Wang et al.]{
 Lin Wang,$^{1,2}$\thanks{E-mail: wl010@bao.ac.cn}
Yiping Shu$^{2}$, Ran Li$^{1,2}$\thanks{E-mail: ranl@bao.ac.cn}, Zheng Zheng$^{1}$, Zhonglue Wen$^{2}$, Guilin Liu$^{3}$
\\
$^{1}$Key laboratory for Computational Astrophysics, National Astronomical Observatories, Chinese Academy of Sciences, Beijing, 100012, China\\
$^{2}$National Astronomical Observatories, Chinese Academy of Sciences, 20A Datun Road, Chaoyang District, Beijing 100012, China\\
$^{3}$CAS Key Laboratory for Research in Galaxies and Cosmology, Department of Astronomy, University of Science and Technology of China, Hefei}

\begin{document}
\label{firstpage}
\pagerange{\pageref{firstpage}--\pageref{lastpage}}
\maketitle

\begin{abstract}
We present Canada-France-Hawaii Telescope (CFHT) MegaCam observations of a galaxy-quasar strong gravitational lens system, SDSS J1640+1932. This system, located at z=0.195 (foreground elliptical galaxy) and z=0.778 (background quasar), was first visually identified by us in the Sloan Digital Sky Survey (SDSS) database. Our CFHT imaging with an angular resolution of 0.7$^{\prime \prime}$ clearly resolves 4 lensed images and a nearly complete Einstein ring. Modeling the system with a singular isothermal ellipsoid (SIE) total mass distribution, we find an Einstein radius of ${2.49^{\prime \prime}}_{-0.049}^{+0.063}$ enclosing a inferred mass of $7.25_{-0.29}^{+0.37}\times10^{11} M_{\odot}$. The quasar and its host galaxy have been magnified by a factor of 23, and the time delay relative to the leading image is determined to be 23.4-25.2 days. These parameters vary minimally when our model is fitted to the {\it{g}}-, {\it{r}}- or {\it{i}}-band images.

\end{abstract}




\section{Introduction}

\label{sec: introduction}
 Strong gravitational lensing is a unique and powerful mass probe. It has been used solely or jointly with other probes to determine the distribution of dark matter \citep[e.g.][]{G07, G08, G09, Nav96, Shu16a, Shu16c}, the total mass density profile and its redshift evolution \citep[e.g.][]{Ko06, Bolton12, Son13, Shu15}, stellar and dark matter decomposition of the lens objects at cosmic distances. The lensing magnification also makes strong lenses natural cosmic telescopes in resolving high-redshift sources. In addition, strong lens systems involving time-variant sources such as quasars measure the Hubble constant \citep{Refs64}, applications of gravitationally lensed quasars such as time delay cosmography \citep[e.g.,][]{Co08, Suyu10} can constrain the quasar luminosity function \citep[]{Rich06} and dark energy \citep[e.g.][]{Koch96, Ogu12}. The first confirmed strong lens system is the Twin Quasar system QSO 0957+561 discovered by \citet{Wal79} which comprises two images of a quasar at redshift 1.41. Since then, several systematic strong lens searches based on photometry and/or spectroscopy have been carried out and led to several hundred of new lenses\citep[e.g.][]{Bolton08, More12, Sone13, Stark13, Shu15, Shu16b, Shu16c, Inada10, Ogu06}. 

In modern cosmology, dark matter shapes the skeleton of the large-scale structure and galactic potential wells, the performing stage for galaxy formation and evolution. Observational characterization of the structure of dark matter haloes imposes key quantitative constraints on current dark matter and galaxy evolution models. Gravitational lensing, especially strong lensing that produces Einstein rings or giant arcs, provides a unique tool for scrutinizing the dark matter halo of galaxies at cosmic distances. \citet{Miralda92} predicted few $\times 10^6$ optical Einstein rings to be detectable over the whole sky. Despite extensive surveys, a limited number of galaxy-scale Einstein rings or giant arcs are known to date \citep[e.g.][]{Leh00,Brow12,Lim09,More12}.

  The SDSS has provided five broad band (u, g, r, i, and z) relatively shallow images with a detection limit of r = 22.5 and a seeing of $1.43^{\prime \prime}$ \citep{Stoughton02}. The huge SDSS data provide an opportunity to systematically search for the rare lensing systems. \citet{Wen09} preformed a search for giant arcs and found 13 new lensing systems by visually inspecting colour images of 39,668 clusters. Another 68 lensing system candidates were found by inspecting colour images of 132684 clusters \citep{Wen11}. Additionally, spectroscopic data of SDSS DR12 was included to search lensing system candidates around the central galaxies of the 132684 clusters. We searched the objects within 10 arcsec of the central
  galaxies, but have a spectroscopic redshift 0.5 greater than that of the central galaxies. Then, we visually inspected their colour images on the SDSS web page\footnote{http://skyserver.sdss.org/dr12/en/tools/chart/list.aspx}. An almost certain lensing system, SDSS\,J164045.66+193257.15 (hereafter SDSS J1640+1932), was found by above procedures.
  
 In this paper, we report the discovery of a galaxy-quasar strong lens system SDSS J1640+1932 identified from visual inspections. Canada-France-Hawaii Telescope (CFHT) follow-up imaging observations in {\it g}, {\it r}, and {\it i} bands show that this system comprises four lensed images of the background quasar and a nearly complete Einstein ring from the quasar's host galaxy. Lens models are further obtained based on the CFHT data, and predictions on the lens galaxy mass within the Einstein radius and time delays between the four lensed images are presented. 
 
This paper is organized as follows: In Section~\ref{sec:strong} we present the CFHT {\it{g}}-, {\it{r}}-, and {\it{i}}-band observations of SDSS J1640+1932. Section~\ref{sec:methods} describes our lens modeling methodology, and Section~\ref{sec:results} presents the results. Finally, discussions and conclusions are presented in Section~\ref{sec:discussions}. For all calculations, we assume a fiducial cosmological model with $\Omega_m = 0.274$, $\Omega_{\Lambda} = 0.726$ and $H_0 \rm = 70\,km\,s^{-1}\,Mpc^{-1}$ \citep[WMAP7;][]{WMAP7}.

\section{Data}
\label{sec:strong}
We identify a new strong quasar lens candidate system SDSS J1640+1932 \footnote{Website:http://skyserver.sdss.org/dr12/en/tools/explore\\/summary.aspx?ra=16:40:45.66\&dec=19:32:57.15} from visual inspection of its SDSS imaging and spectroscopic data. The left panel in Fig.1 shows the SDSS image (gri colour composite) centred on an early-type galaxy SDSS J1640+1932. Bluish features $\approx 2.5\arcsec$ away from the central galaxy are clearly seen and identified by SDSS as a single object SDSSJ164045.57+193255.27. Furthermore, another fainter clump is present to the north of the central galaxy. This image configuration suggests this system to be a strong lens candidate. Both objects are observed spectroscopically by the BOSS survey, and their spectra are shown on the right panels in Fig.1. The redshifts of both objects are thus determined to be 0.195 and 0.778 respectively.\par 
 
We hereby proposed CFHT/MegaCam follow-up campaign to obtain higher-resolution imaging of this candidate: SDSS J1640+1932. We imaged it in the {\it{g}}-, {\it{r}}- and {\it{i}}-band with the CFHT/Megacam on 2015 September 10th. The wide-field imager, MegaCam (built by CEA, Saclay, France), consists of 40 $2112\times4644$ pixel CCDs, covering a full $1\times1$ square degree field-of-view with a resolution of 0.185 arcsecond per pixel to properly sample the 0.7 arcsecond median seeing offered by the CFHT at Mauna Kea.

The spectrum of the central galaxy has a typical spectral shape of an elliptical galaxy, and its redshift is found to be 0.195. The spectrum of the extended arc has strong atomic emission lines, and suggests it to be a quasar at z=0.778. We co-added the 8 exposures in {\it{g}}-band(135 seconds each), 7 exposures in {\it{r}}-band (280 seconds each) and {\it{i}}-band (140 seconds each) using SExtractor, SCAMP and SWarp\footnote{Website: http://www.astromatic.net/software/}, respectively. Firstly, we use SExtractor to build a catalogue of objects from every image and calculate statistic error on each pixel. Next, we use SCAMP to read the catalogs and compute astrometric and photometric solutions. Finally, we employ SWarp to resample and co-add together these 8 images to create a long-time exposure image. The central 61$\times$61 pixels cutout containing only the lens and images of the source is extracted for further analyses. Fig.~\ref{Fig2} shows the final image after co-adding. {It clearly shows a central galaxy surrounded by an extend blueish arc in the southwest (labeled as C and D) and two other blobs in the north (A) and southeast (B), respectively. Both the central galaxy and the extended arc are observed spectroscopically by the SDSS, and their spectra are shown in the right panel of Fig.1. With the aid of the better imaging resolution provided by the CFHT/Megacam, the blueish arc is resolved into two blobs (labeled as C and D). Furthermore, another faint blob in the southeast (B) and a red, almost full Einstein ring are clearly detected for the first time.} This is certainly a strong lensing system of quadruple images, the background galaxy contains a quasar which is very special because we can see not only the quasar but also its host galaxy.

\begin{figure*}
	\begin{center}
		\resizebox{135mm}{!}{\includegraphics{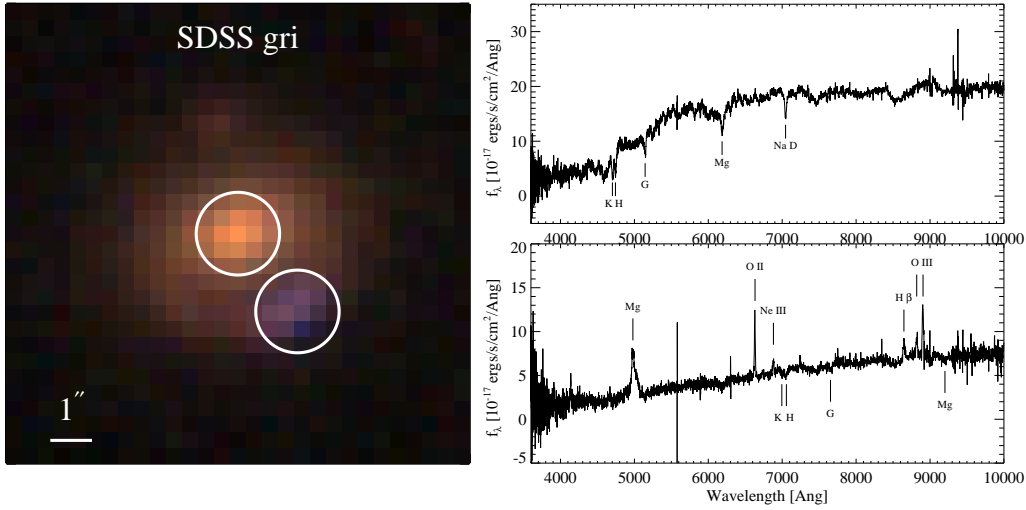}}
	\end{center}
	{\caption{Left panel: The SDSS image of the lens system. The fiber subtend an angle of 2 arcsecond (white circle). Right panels: The SDSS-measured spectra of the lens galaxy (top) and the lensed quasar (bottom). 
}}\label{Fig1}
		
\end{figure*}

\begin{figure}
	\begin{center}
		\resizebox{60mm}{!}{\includegraphics{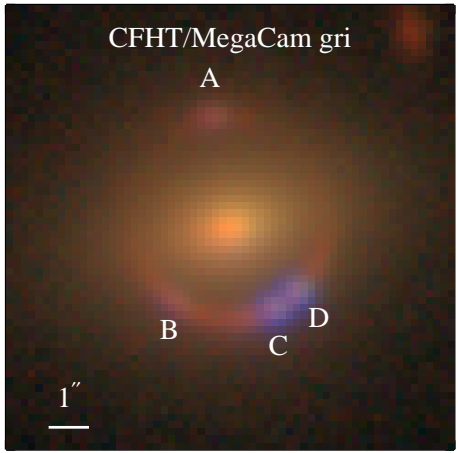}}
	\end{center}
	{\caption{The coadded data of {\it{g}}-band, {\it{r}}-band and {\it{i}}-band taken by CFHT}}\label{Fig2}
\end{figure}

\section{Lens Modeling Strategy}\label{sec:methods}

 The system is modeled by the lens modeling tool {\tt lfit\_gui} developed by \citet{Shu16c}. Here we briefly describe the settings used in this work. 

\subsection{Foreground-light and Source Models}

We model the surface brightness distribution of the lens galaxy with on elliptical S\'{e}rsic profile defined as 
\begin{equation}
I=I_{e} \exp{\{-b(n)[(\frac{r}{r_{e}})^\frac{1}{n}-1]\}}
\end{equation}
in which I$_e$ is the surface brightness at the effective radius r$_e$ that encloses half of the total light. The constant b(n) is defined in \citep{Cio99, Mac03}\footnote{Website: http://www.slac.stanford.edu/$\sim$oguri/glafic/}.

 As the CFHT data clearly resolve the Einstein ring formed from the quasar's host galaxy, we model the source surface brightness distribution with one elliptical S\'{e}rsic model. We choose not to model the quasar itself because the Einstein ring already provides robust constraints on the lens model.

\subsection{Lens model}
{ 
 The projected surface-mass distribution of lens galaxy is modeled by singular isothermal ellipsoid profile (hereafter SIE) (for details, see \citealt{Kor94}). Let $\bm{\eta}$\ denote the two-dimensional position of source on the source plane, $\bm{\xi}$ the impact vector of light ray in the lens plane, both measured with respect to an `optical axis'. The gravitational lensing equation shows as
       \begin{equation}
       \bm{\eta}= \frac{D_s}{D_d}\bm{\xi} - D_{ds}\hat{\bm{\alpha}}(\bm{\xi}),
       \label{eq4}
       \end{equation}      
where $D_d, D_s, D_{ds}$ denote the angular diameter distance to the lens, the source, and from the lens to the source, respectively. $\hat{\bm{\alpha}}(\bm{\xi})$ is the deflection angle, related to the surface mass density $\Sigma(\bm{\xi})$ by
		\begin{equation}
		\hat{\bm{\alpha}}(\bm{\xi}) = \frac{4G}{c^2}\int_{\bm{R^2}} \Sigma(\bm{\xi}')\frac{\bm{\xi}-\bm{\xi}'}{|\bm{\xi}-\bm{\xi}'|^2}\,d^2{\xi}'.
		\end{equation}					
Introducing angular coordinates by
         \begin{equation}
         \bm{\eta}= D_{s}\bm{\beta} \quad \mathrm{and} \quad \bm{\xi} = D_{d}\bm{\theta}, 
         \end{equation}
we can transform Eq.(\ref{eq4}) to
           \begin{equation}            	
   \bm{\beta} = \bm{\theta} - \frac{D_{ds}}{D_{s}}\hat{\bm{\alpha}}(D_{d}\bm{\theta}) \equiv \bm{\theta} - \bm{\alpha}(\bm{\theta}),
            \label{eq5}	          	
          	\end{equation}
the interpretation of the lens Eq.(\ref{eq5}) is that a source with true position $\bm{\beta}$ can be seen by an observer to be located at angular positions $\bm{\theta}$ satisfying Eq.(\ref{eq5}). So the dimensionless lens equation is obtained as
        \begin{equation}
        \bm{y}=\bm{x}-{\bm{\alpha}}(\bm{x}),
        \end{equation}
where $\bm{x}$, $\bm{y}$ and length scale ${\xi_0}$ are defined as
        \begin{equation}
\bm{x}=\frac{\bm{\xi}}{\xi_0},  \qquad  \bm{y}=\frac{\bm{\eta}}{\eta_0}
\quad \mathrm{with} \quad\eta_0 = \frac{D_s}{D_d}\xi_0 ,
          \end{equation}
          
        \begin{equation}
        \xi_0=4\pi\left(\frac{\sigma_{SIE}}{c}\right)^2\frac{D_d D_{ds}}{D_s} 
        \end{equation}
 and ${\sigma_{SIE}}$ is the lensing velocity dispersion. 
     
 We can express the scaled deflection angle in terms of the surface mass density as
	\begin{equation}
	\bm{\alpha}(\bm{x})=\frac{1}{\pi}\int_{\bm{R^2}} {\kappa}(\bm{x}')\frac{\bm{x}-\bm{x}'}{|\bm{x}-\bm{x}'|^2}\,d^2x',
	\end{equation}
where we have defined the dimensionless surface mass density or convergence
	\begin{equation}
	{\kappa}(\bm{x})=\frac{\Sigma(\xi_0\bm{x})}{\Sigma_{cr}}\quad \mathrm{with} \quad\Sigma_{cr}=\frac{c^2D_s}{4\pi GD_dD_{ds}}.
	\end{equation}

As for the SIE profile, the lens equation can be read as
	\begin{equation}
	\bm{y}=\bm{x}-\frac{\sqrt{f}}{f'}\left[\mathrm{arcsinh}\left(\frac{f'}{f}\cos\varphi\right)\bm{e_1}+\arcsin\left(f' \sin\varphi\right)\bm{e_2}\right]
	\end{equation}
	where $\bm{e_i}$ is the unit vector in the direction of $x_i$, f is the axis ratio which always taken as 0$ < f \leq $1 and $\varphi$ the position angle of vector $\bm{x}$ relative to elliptical's major axis.\par
	As a consequence, the expressions for the magnification\citep[e.g.][]{Kor94}
	\begin{equation}
	\mu=\frac{1}{(1-2\kappa)}
	\label{magnification}
	\end{equation}
	We also introduce the deflection potential
	\begin{equation}
	\Psi(\bm{x})=\frac{1}{\pi}\int_{\bm{R^2}} \bm{\kappa}(\bm{x}')\ln|\bm{x}-\bm{x}'|d^2x'\quad
	\end{equation}
	which is related to the deflection angle by
	\begin{equation}
	\bm{\alpha}(\bm{x})=\nabla\Psi(\bm{x})
	\end{equation}
	and the time delay \citep{Nara96} can be written
	\begin{equation}
	t\bm{(\theta)}=\frac{1+z_{d}}{c}\frac{{D_{d}D_{s}}}{D_{ds}}
	\left[\frac{1}{2}(\bm{\theta}-\bm{\beta})^2-\Psi(\bm{\theta})\right]
	\label{time}
	\end{equation}
	where $\bm{\beta}$ and $\bm{\theta}$ are the angular separations of the source and the image from the optic axis as seen by the observer, respectively.\par

\subsection{Optimization}

 As discussed in \citet{Bolton06}, \citet{Marshall07}, and \citet{Shu16a}, we fit the foreground light and mass simultaneously in order to reduce the systematic errors. The foreground-light distribution model is combined with the predicted lensed images and convolved with PSF. The combined model is then compared to the observed data, and the parameter optimization is done by minimizing a $\chi^2$ function which is defined as
	   \begin{equation}
    \chi^2 = \sum_{i,j}\left[\frac{I_{i,j}^{\mathrm{data}} - \left(I_{i,j}^{\mathrm{fg}} + I_{i,j}^{\mathrm{image}}\right) }{\sigma_{i,j}}\right]^2
     \end{equation}	
where $I_{i,j}^{\mathrm{data}}$, $I_{i,j}^{\mathrm{fg}}$, $I_{i,j}^{\mathrm{image}}$ are the observed, PSF-convolved foreground lens, and PSF-convolved lensed image intensities at pixel ($\emph{i, j})$, in the image plane, respectively. $\sigma_{i,j}$ is the corresponding rescaled pixel count error. All the model parameters are optimized using the Levenberg-Marquardt algorithm with the LMFIT package \citep{Newville14} and the parameter uncertainties are estimated by MCMC approach.		
	
\section{RESULTS}\label{sec:results}
	
 Among the observed CFHT three band images, the Einstein ring appears the least prominent in the {\it{g}} band, while the exposure time for the {\it{r}} band is the longest, resulting in the highest available signal-to-noise ratio. Therefore, in this section, we focus on reporting our analysis results from the {\it{r}}-band data, which provides more robust constraints on the lens model. 
	
Fig.~\ref{Fig3} shows the performance of the foreground-light subtraction with the original data, foreground-light model, and the foreground-subtracted residual displayed from left to right, respectively. It can be seen that the one-S\'{e}rsic model provides a satisfactory fit to the lens galaxy light distribution. The best-fit parameters are summarized in Table~\ref{table1}. The morphological parameters of the lens galaxy are consistent across three filter bands. 

In Fig.~\ref{Fig4}, we show the foreground-subtracted residual followed by the best-fit lensed image model, the final residual, and the source model. The white lines in the first three panels represent the critical line and the white line in the last panel is the caustic. The system is in the so-called 'fold' configuration in which the source is on the edge of the caustic and two of the four images are going to merge. We can see that the one-SIE plus one-source model is able to recover the majority of the lensed features, especially the Einstein ring. We further investigate the need of extra external shear, but find it to be negligible. The best-fit parameters of the mass model based on the three bands data are summarized in in Table ~\ref{table3}.
	
The Einstein radius $\theta_{E}$ is estimated to be ${2.49^{\prime \prime}}_{-0.049}^{+0.063}$, and the total enclosed mass within the Einstein radius is $7.25_{-0.29}^{+0.37}\times10^{11} M_{\odot}$. The characteristic lensing velocity dispersion $\sigma_{SIE}$ is related to $\theta_{E}$ as
	\begin{equation}
	\theta_{E}=4\pi\frac{\sigma_{SIE}^{2}}{c^{2}}\frac{D_{ds}}{D_{s}},
	\end{equation}
and is estimated to be 349 km/s. It is consistent with the measured stellar velocity dispersion of 379 km/s for SDSS J1640+1932 based on single-fiber spectroscopic data. The average magnification, defined as the ratio between the observed total light of the lensed images and that of the source, is determined to be 23. The time delays between different lensed images can be obtained through Equation Eq.(\ref{time}). Using our best-fit results, we have derived the time delays and magnitudes for A, B, C and D (see Table ~\ref{table2}). Compared to other strong lensing systems containing gravitationally lensed quasars, these results render typical values.
	
For completeness and cross-checking purposes, we have also performed the fitting on {\it{g}}- and {\it{i}}-band as well, finding minimal difference from the {\it{r}}-band results (see Table ~\ref{table3} for details).

\begin{table*}
	\centering
	\renewcommand\arraystretch{1.1}
	\caption[]{The parameters of {S\'{e}rsic} model about luminosity distribution of the lens galaxy  }
	\begin{tabular}{p{2.8cm}|p{2.3cm}|p{2.8cm}|p{2.9cm}|p{2.9cm}|p{2.9cm}}
		\hline
		\multicolumn{2}{c}{Parameters} & \multicolumn{3}{c}{ Best-Fit Values } \\
		\hline
		\multicolumn{2}{c}{} &   \it{g} &   \it{r} &   \it{i}\\
		\hline

		Offset in RA & $\Delta \alpha \ ('' )$ & $0.052_{-0.0046}^{+0.0045} $& $0.060_{-0.0065}^{+0.0067} $& $0.0_{-0.00020}^{+0.00020} $  \\
		Offset in Dec & $\Delta \delta \ ('')$ & $-0.030_{-0.0029}^{+0.0029}$ & $0.0050_{-0.00060}^{+0.00056}$& $-0.0_{-0.00030}^{+0.00030}$  \\
		Effective radius& r$_{s} \ ('' )$ & $2.37_{-0.19}^{+0.20}$ & $1.80_{-0.23}^{+0.26}$& $2.15_{-0.22}^{+0.22}$ \\
		Axial Ratio& q & $0.64 _{-0.055}^{+0.042}$ & $0.57 _{-0.069}^{+0.069}$& $0.60 _{-0.070}^{+0.080}$\\
		Position angle  & $\phi$ \ (deg) & $ 102.59_{-8.56}^{+8.66}$& $ 102.83_{-10.14}^{+11.23}$& $ 102.83_{-11.72}^{+12.14}$ \\
		Power index &n &$ 5.62 _{-0.49}^{+0.45}$&$ 5.91 _{-0.69}^{+0.79}$&$ 3.33 _{-0.39}^{+0.42}$\\
		
		\hline
	\end{tabular}
	
	\footnotemark{ -All angular offsets are with respect to $\alpha$=16$^h$40$^m$45.$^s$57, $\delta$=19$^{\circ}$32$^{'}$55.$^{''}$27 (J2000).}
	\label{table1}
\end{table*}

\begin{table}
	\begin{center}
		\caption{\label{table2}The time delays, the magnitudes in {\it{r}}-band and the {\it g-r} colour about the four images }
		\begin{tabular}{l  c c c c}
			\hline \hline
			{images}&time delay (days) & magnitudes & g-r  \\
			\hline
		    {   A  }  &   ----   & {22.11 (A)}& {0.86 (A)} \\
			{	B-A} & {25.23} & {20.98 (B)} & {0.68 (B)} \\
			{	C-A} & {23.54} & {19.97 (C)} & {0.35 (C)}\\
			{	D-A} & {23.43} & {20.35 (D)}& {0.38 (D)} \\
			\hline
		\end{tabular}
	\end{center}
\end{table}

\begin{figure*}
	\begin{center}
		\includegraphics[width=16.1cm,height=5.1cm]{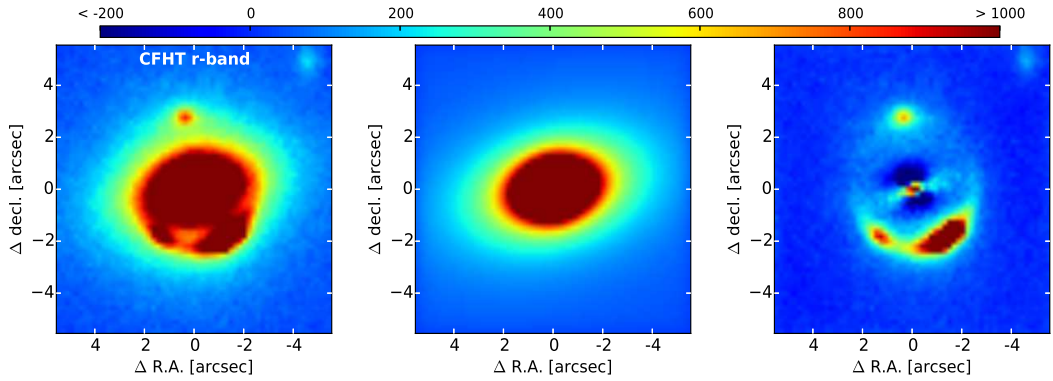}
	\end{center}
	\caption{The presentation of the foreground-light subtraction in {\it{r}}-band. From left to right, the first image is the original data, then the photometric fitting image and the last is the source background galaxy.
		\label{Fig3}}
\end{figure*}

\begin{table*}
	\centering
	\renewcommand\arraystretch{0.88}
	\caption{Lens modeling parameters and results
		\label{table3}}
	\begin{tabular}{p{2.cm}|p{2.2cm}|p{2.cm}|p{2.cm}|p{2.cm}}
		\hline
		
		\multicolumn{2}{c}{Parameters} & \multicolumn{3}{c}{ Best-Fit Values } \\
		\hline
		\multicolumn{5}{c|}{Lens}  \\
		\hline
		\multicolumn{2}{c}{} & \it{g} & \it{r} & \it{i}\\
		\hline
		Einstein radius &$\theta_{E}$ & $2.48_{-0.15}^{+0.14}$& $2.49_{-0.049}^{+0.063} $& $2.50_{-0.11}^{+0.10} $\\
		Offset in RA & $\Delta \alpha_{lens} \ ('' )$ & $0.054_{-0.0038}^{+0.0037} $ & $0.063_{-0.0091}^{+0.0071} $& $0.060_{-0.0074}^{+0.0060} $\\
		Offset in Dec & $\Delta \delta_{lens}\ ('' )$ &$ 0.078 _{-0.0068}^{+0.0066}$& $0.11_{-0.013}^{+0.013} $& $0.11_{-0.014}^{+0.011} $ \\
		Axial Ratio& q$_{lens}$ &$ 0.87_{-0.039}^{+0.041}$& $0.89_{-0.17}^{+0.19} $& $0.88_{-0.017}^{+0.018} $ \\
		Position angle  & $\phi$$_{lens}$ \ (deg) &$ 99.74_{-7.70}^{+5.99}$& $100.97_{-8.80}^{+9.03} $& $100.22_{-10.11}^{+8.04} $ \\
		\hline
		\multicolumn{5}{c|}{Source}\\
		\hline
		Offset in RA & $\Delta \alpha_{s} \ ('' )$ & $0.038_{-0.0029}^{+0.0029}$ & $0.043_{-0.0052}^{+0.0057} $& $0.039_{-0.0052}^{+0.0052} $ \\
		Offset in Dec & $\Delta \delta_{s}\ ('' )$ & $0.25_{-0.022}^{+0.019}$ & $0.26_{-0.033}^{+0.025} $& $0.27_{-0.021}^{+0.023} $\\
		Axial Ratio& q$_{s}$ &$ 0.18 _{-0.013}^{+0.020}$& $0.38_{-0.051}^{+0.065} $& $0.38_{-0.033}^{+0.034} $ \\
		Position angle  & $\phi_{s}$ \ (deg) & $129.90_{-11.61}^{+9.75} $& $100.04_{-12.43}^{+10.91} $& $94.50_{-9.78}^{+9.34} $\\
		Power index &n & $1.14_{-0.26}^{+0.22} $& $7.83_{-0.29}^{+0.39} $& $6.61_{-0.24}^{+0.33} $\\
		Effective radius& r$_{s}$& $0.021_{-0.0013}^{+0.0016} $& $2.39_{-0.20}^{+0.18} $& $2.19_{-0.26}^{+0.24} $\\
		\hline
	
	\end{tabular}
	
	\footnotemark{ -All angular offsets are with respect to $\alpha$=16$^h$40$^m$45.$^s$57, $\delta$=19$^{\circ}$32$^{'}$55.$^{''}$27 (J2000).}
	
\end{table*}

\begin{figure*}
	\begin{center}
		\resizebox{180mm}{!}{\includegraphics{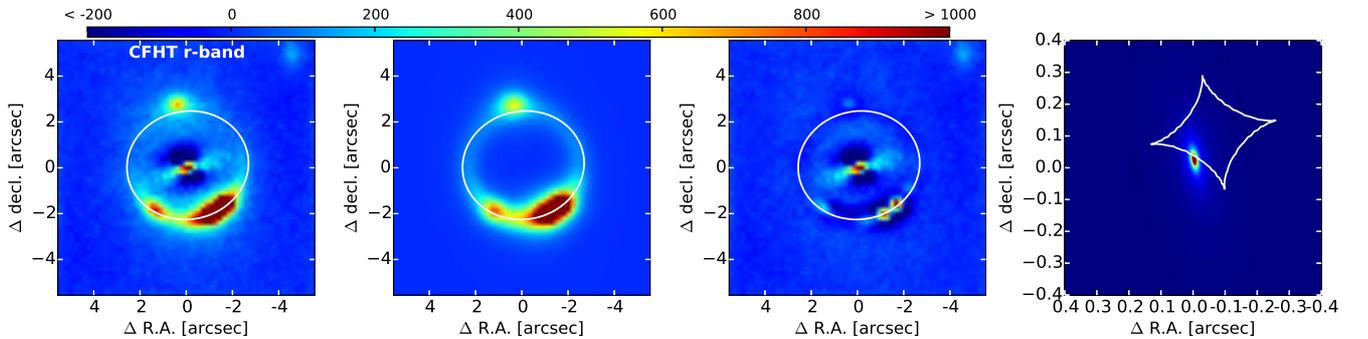}}
	\end{center}
	\caption{The reconstruction of mass model in {\it{r}}-band. From left to right, the first is the source background galaxy, the best-fit mass model of source background galaxy, the residual image of the source galaxy subtracted by the mass model and the last is the image of the source. The white elliptical line represents critical line and the caustic are also plotted in the image of the source.  
		\label{Fig4}}
\end{figure*}

\begin{figure}
	\begin{center}
		\resizebox{86mm}{!}{\includegraphics{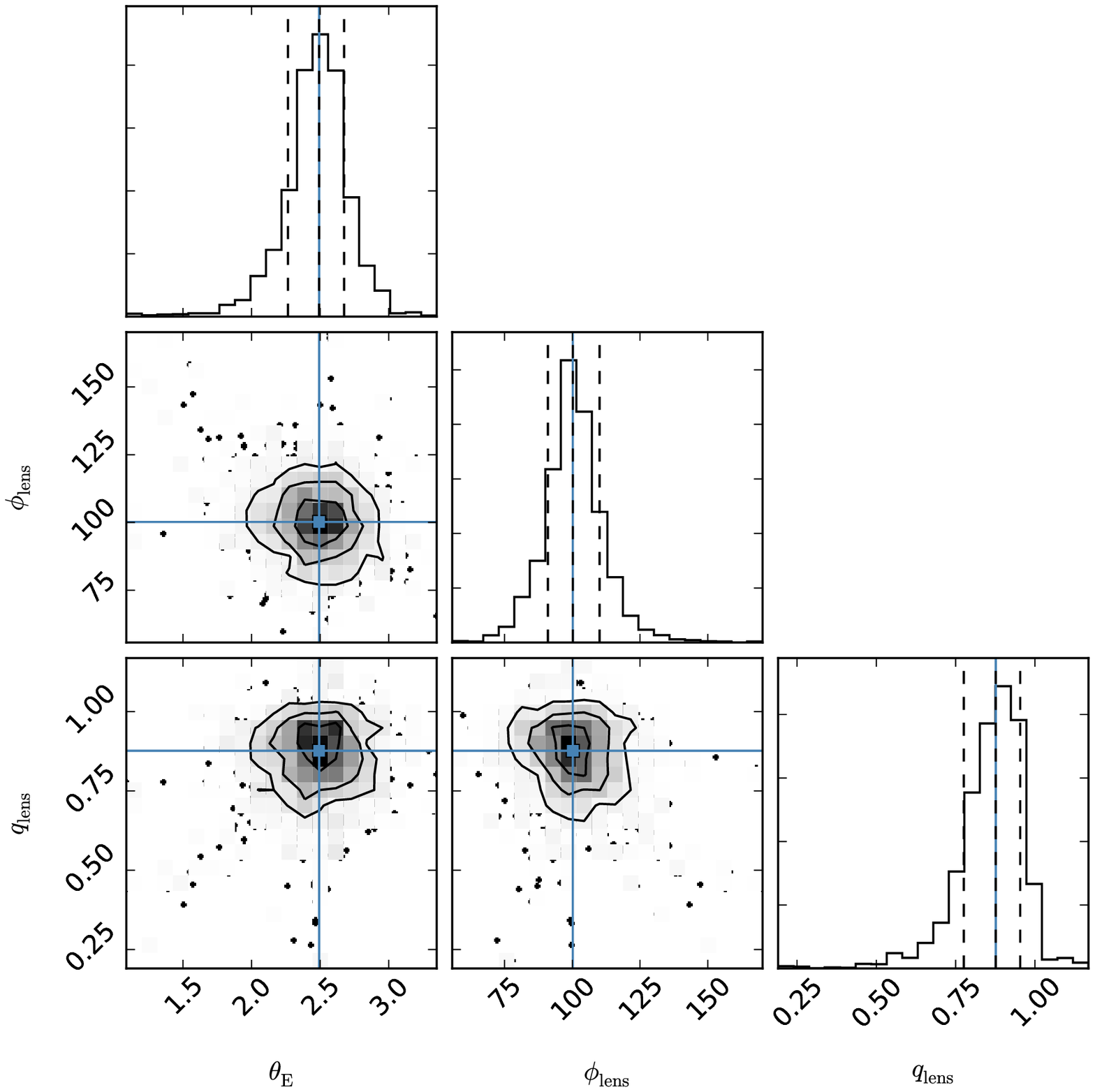}}
	\end{center}
	\caption{Reconstructed lens model about {\it{r}}-band. Inference on the three main parameters of the lens model. The distribution of axial Ratio q$_{lens}$, Position angle $\phi$$_{lens}$ and Einstein radius $\theta_{E}$ about the lens galaxy. For each of the posterior distributions, we mark the 16\%, 50\% and 84\% condence interval (vertical dashed lines). 
		\label{param}}
\end{figure}

\section{DISCUSSIONS AND CONCLUSIONS}\label{sec:discussions}

SDSS J1640+1932 is a special galaxy-quasar strong lens system because the quasar's host galaxy is lensed into a clearly visible, nearly complete Einstein ring. For most lensed quasars, their hosts are outshone by the much brighter central quasars. For instance, out of the 28 lensed quasars and candidates in the SDSS Quasar Lens Search \citep{Ogu06, Inada12}, only five systems show clearly detected host galaxies as recently revealed by the Subaru adaptive optics observations \citep{Rusu16}. Detections of Einstein rings formed by the host galaxies are even rarer. Systems like SDSS J1640+1932 can play an important role in the determination of the Hubble constant from time-delay cosmography. One of the systematics in this method is the lens mass model. Extra constraints from the Einstein rings will allow much more accurate mass-distribution reconstructions, and therefore largely reduce the systematics.

Two of the four lensed images (images C and D) appear much bluer than the other two images (images A and B). Quantitively, the {\it g-r} colours of images A, B, C and D are 0.86, 0.68, 0.35 and 0.38, respectively. We propose three possible interpretations. 
\begin{itemize}
		
{\item Differential magnifications between the bluer quasar and the redder host galaxy. The quasar can be considered as a point source, while the light distribution of its host galaxy is much more extended and smoother. It is known that the magnification changes rapidly near the critical curve. Images C and D are almost on the critical curve, and hence magnifications on these positions are much higher than at images A and B. As a result, the light at positions of images C and D is dominated by the highly magnified quasar light, which is bluer than its host galaxy. As for images A and B, the quasar is less magnified. The finite image resolution will further make the quasar light smear out. The net effect is that the colours at images A and B are dominated by the quasar's host galaxy. 

\item Differential dust extinction. If images C and D suffer less from the dust extinction, then they could appear bluer. But this scenario is not very likely. The lens galaxy is an elliptical galaxy which should contain little dust. In addition,  the four images locate at about the same distances from the lens galaxy, so significant differential dust extinction is not expected. }

\item Time delays. According to our lens model, light from images C and D arrives at about the same time, but later than image A by $\sim 23$ days and earlier than image B by $\sim 2$ days. If the brightness of the quasar happens to increase dramatically (by at least 0.5 mag) when light from images C and D arrives and drops back after 2 days, then images C and D could be bluer as observed. However, such a possibility would be very low. First of all, similar colour differences were already seen in the SDSS image which was taken about $\sim 10$ years before our CFHT observations. The chance of catching similar variabilities in two totally uncorrelated observations is negligible. Furthermore, light-curve observations for some known lensed quasars suggest that the typical magnitude variation for a quasar is 0.3-0.8 mag over 100-200 days \citep[e.g.,][]{Eulaers13, Tewes13, RathnaKumar13}. 
 
\end{itemize}\par

 In this paper, we present follow-up CFHT multi-band imaging observations of a new galaxy-quasar strong lens system SDSS J1640+1932 identified based on the SDSS imaging and spectroscopic data. The lens galaxy is located at $z=0.195$ while the background quasar is at $z=0.778$. The high-resolution CFHT data clearly resolve the four lensed quasar images and a nearly complete Einstein ring from the quasar's host galaxy. By modeling the total mass distribution of the lens galaxy as an SIE profile, we infer that the Einstein radius of this system is ${2.49^{\prime \prime}}_{-0.049}^{+0.063}$, and the total enclosed mass within the Einstein radius is $7.25_{-0.29}^{+0.37}\times10^{11} M_{\odot}$. The lensing velocity dispersion is consistent with the measured stellar velocity dispersion, and external shear is tested to be negligible. The quasar's host galaxy can be well modeled by a single S\'{e}rsic component. The source is highly magnified with an average magnification of 23 in the CFHT {\it r}-band. Time delays between different lensed images are predicted. Robust constraints on the lens galaxy mass distribution provided by the Einstein ring make this system a potential candidate for time-delay cosmography study. We plan to carry out follow-up observations on the other lens candidates found in \citet{Wen11} and expect to discover more new strong lens systems. \par

\section*{Acknowledgements}

 We thank the anonymous referee for very constructive comments that help to improve the quality of this paper significantly. The work is based on observations obtained with MegaPrime/MegaCam, a joint project of CFHT and CEA/DAPNIA, at the Canada-France-Hawaii Telescope (CFHT) which is operated by the Nationa Research Council (NRC) of Canada, the Institut National des Science del'Univers of the Centre National de la Recherche Scientifique (CNRS) of France, and the University of Hawaii. We thank for Telescope Access Program (TAP) which is a program to give astronomers based in China direct access to competitive instrumentation on intermediate- and large-aperture optical/infrared telescopes. We acknowledges NSFC grant (Nos.11303033,1151113005411333001,11133003, 11425312).
RL is supported by Youth Innovation Promotion Association of CAS. Y.S. is partially supported by the National Natural Science Foundation of China (NSFC) grant 11603032. Guilin Liu is supported by the National Thousand Young Talents Program of China, and acknowledges the grant from the National Natural Science Foundation of China (No. 11673020 and No. 11421303) and the Ministry of Science and Technology of China (National Key Program for Science and Technology Research and Development, No. 2016YFA0400700).





\bsp	
\label{lastpage}

\end{document}